\documentclass[conference]{IEEEtran}

\usepackage{graphicx}
\usepackage{amsmath}
\usepackage{listings}
\usepackage{color}

\definecolor{mygreen}{rgb}{0,0.6,0}
\definecolor{mymauve}{rgb}{0.58,0,0.82}

\begin{document}

\title{A Bespoke Workflow Management System for Data-Driven Urgent HPC}

\author{\IEEEauthorblockN{Gordon P S Gibb, Nick Brown, \\ Rupert W Nash}
\IEEEauthorblockA{EPCC \\
The University of Edinburgh\\
Bayes Centre\\
47 Potterrow\\
Edinburgh, EH8 9BT, UK\\
Email: g.gibb@epcc.ed.ac.uk}
\and
\IEEEauthorblockN{Miguel Mendes, Santiago Monedero,\\ Humberto Díaz Fidalgo, \\Joaquín Ramírez Cisneros, \\ Adrián Cardil}
\IEEEauthorblockA{Edificio Tecnosylva\\
Parque Tecnológico de León\\
24009 León\\
Spain}
\and
\IEEEauthorblockN{Max Kontak}
\IEEEauthorblockA{DLR German Aerospace Center\\
Institute for Software Technology\\
High-Performance Computing\\
Linder Höhe\\
51147 Köln\\
Germany}
}


%


\maketitle

\begin{abstract}
In this paper we present a workflow management system which permits the kinds of data-driven workflows required by urgent computing, namely where new data is integrated into the workflow as a disaster progresses in order refine the predictions as time goes on. This allows the workflow to adapt to new data at runtime, a capability that most workflow management systems do not possess. The workflow management system was developed for the EU-funded VESTEC project, which aims to fuse HPC with real-time data for supporting urgent decision making.  We first describe an example workflow from the VESTEC project, and show why existing workflow technologies do not meet the needs of the project. We then go on to present the design of our Workflow Management System, describe how it is implemented into the VESTEC system, and provide an example of the workflow system in use for a test case.
\end{abstract}


%
\IEEEpeerreviewmaketitle

\section{Introduction} \label{introduction}

Workflows are ubiquitous in science, technology and business. Simply put, a workflow is the collection of tasks and their ordering, required to achieve an objective. The compilation process for a piece of software is a good example of a computing workflow, where a number of files need to be compiled in a specific order according to their dependencies and then finally linked together to form the executable. Often a workflow can be visualised as a flowchart or directed graph, with the nodes being each individual task that needs to be carried out, and the connections between nodes representing the order the tasks must be conducted.

Whilst a workflow can be carried out manually, it is often more convenient or necessary to automate its execution. The simplest case of this would be via a simple script with the stages written in order. Continuing with the aforementioned compilation example, this would be akin to writing a bash script containing each compilation step in order. This approach is acceptable for a reasonably small and simple workflow, although becomes cumbersome for large and/or complex workflows. Additionally, if multiple workflows are required, a bespoke script must be written for each individual workflow which may not scale well for large numbers of workflows. For such cases, a dedicated {Workflow Management System} (WMS) is desirable. With such systems the tasks in a workflow are defined along with their dependencies on other tasks. The WMS determines the order these need to be executed in, and executes them in the correct order. An analogy can be drawn to writing a \emph{Makefile} which contains the description of the compile tasks and their dependencies. This is then executed by running the \emph{make} command (effectively the WMS) to undertake the actual compilation. Multiple executables can be compiled using the same \emph{make} approach, with the user only needing to write an individual configuration (e.g. a makefile) for each one.

There are numerous WMSs, such as Apache Taverna \cite{taverna}, Luigi \cite{luigi}, Snakemake \cite{snakemake}, Toil \cite{toil} and Drake \cite{drake}. Whilst these all have different interfaces and are designed for specific types of workflow, the core functionality remains the same. In an attempt to standardise workflow descriptions, the Common Workflow Language (CWL) \cite{cwl,cwlweb} was created. This  sets out how workflows are described, using YAML files to describe the tasks and dependencies between them. Many WMSs have been (or are being) expanded to allow their workflows to be described using CWL. However, traditionally many such workflow systems tend to be limited to workflows that have limited to no conditional branching, as they are unable to determine the runtime order for such steps.

Workflows are an essential part of urgent computing \cite{brown2019role}, and in order to optimise the speed of the response these workflows should be automated as much as possible so as to reduce the slowdown introduced by human reaction time. WMSs are therefore ideal candidates to automate urgent computing workflows. The EU funded  Visual  Exploration  and  Sampling  Toolkit  for  Extreme Computing (VESTEC) project \cite{vestec} aims to fuse HPC with real-time data to prove its effectiveness in supporting urgent decision making. In order to demonstrate this, the project is based around three use cases: wildfires, mosquito borne diseases and space weather. For each, forecasts of the developing disaster will be produced, and constantly refined and updated by new data as it becomes available. Each use case will have a workflow associated with it, whereby data must be acquired by various means and numerous simulations must be run in order to produce forecasts of the disaster's progression, which can be fed to the urgent decision maker to help with planning mitigating actions to be taken. 

VESTEC's design architecture, covered in detail in \cite{vestec_system}, consists of a centralised control system (running on a server), which has access to multiple HPC machines for running simulations, a web API to allow urgent decision makers to connect to the system, and functionality for acquiring real-time data from sources such as satellites and sensors. The control system adopts a modular approach, whereby each function it has to perform is represented by a single component, and the components work together to make up the full control system. One such vital component is the workflow manager, which actually executes the workflows (e.g. one of the three VESTEC use cases), incorporating data acquired from satellites, and any input from the urgent decision maker.

In this paper, we describe the workflow manager component we have developed for the VESTEC system. In Section \ref{wildfire_sec} we describe the wildfire use case's workflow in detail in order to illustrate the requirements for a WMS, and demonstrate why these requirements preclude existing workflow technologies. In Section \ref{design_sec} we establish a list of requirements for our workflow manager, and discuss some of the design features it requires. The implementation of the workflow manager is then described in Section \ref{implementation_sec}, and its integration with the rest of the VESTEC system is covered in \ref{vestec_integration}. Finally, a summary of the paper and our conclusions can be found in Section \ref{conclusion_sec}.

\begin{figure*}[t]
\centering
\includegraphics[width=\textwidth]{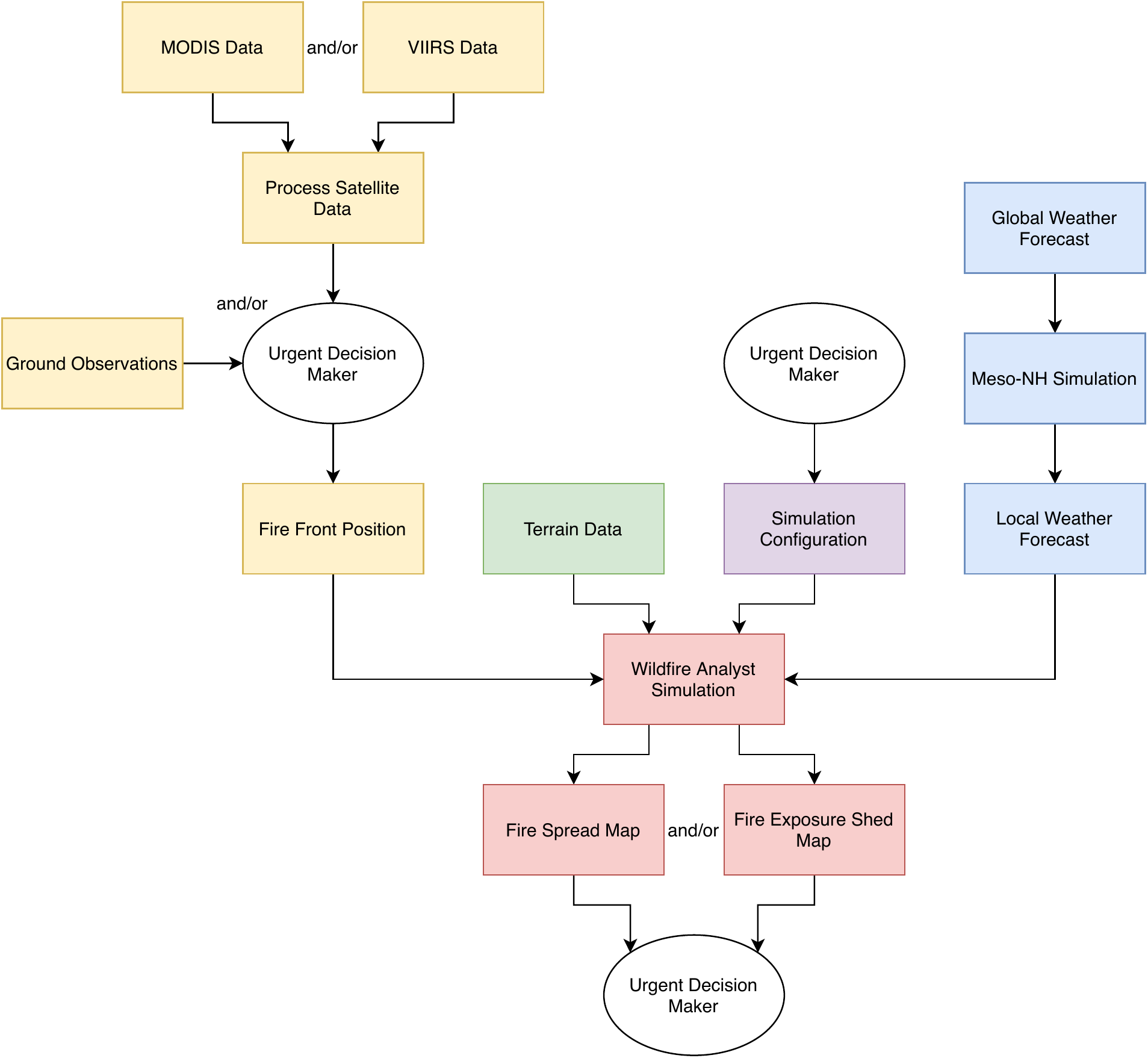}
\caption{The wildfire workflow as a flowchart. Wildfire Analyst requires four inputs and can produce two different kinds of outputs depending upon the configuration file provided to it. Several steps are required to provide the weather and fire position requirements, and for the latter there are several possible ways to obtain the input data. The urgent decision maker is required to specify the simulation configuration along with the fire front position. The colours group the different sub-workflows.}
\label{wildfire_flowchart}
\end{figure*}

\section{Workflow example: Fighting wildfires} \label{wildfire_sec}

In order to provide context and motivation for the kinds of workflows the VESTEC system needs to execute, it is prudent to study the workflow required for one of our use cases. To this end we present the wildfire use case. The aim of this is to predict the spread of a wildfire so that urgent decision makers can decide how to best allocate resources to tackle the fire and determine if evacuation is necessary. As the fire progresses and new data becomes available, this data can be used to refine the forecasts. The fire is simulated over a spatial Area Of Interest (AOI) defined by the urgent decision maker. A flowchart outlining the full workflow is shown in Figure \ref{wildfire_flowchart}.

The simulation code used to run the fire simulations is called Wildfire Analyst (WFA), developed by Tecnosylva \cite{wildfire_analyst,wildfire_analyst_web}. WFA allows for new data to be pushed to it whilst it is running, and it uses this to update and improve its predictions. Wildfire Analyst can produce two types of forecast: a forward in time simulation of the spread of the fire perimeter, or for a given time the probabilistic exposure shed (e.g. the likelihood of the fire having reached a given point by a certain time) for the AOI. The kind of forecast produced is specified in a configuration file supplied to WFA. In addition to the configuration file, Wildfire Analyst requires three inputs which will be described in the following paragraphs.

Firstly, a description of the location of the fire at an initial time is required. This could be the ignition point of the fire (if known), or the known position of the fire front at a certain time. This information can be obtained from observers on the ground, or from satellite sources such as the Moderate Resolution Imaging Spectroradiometer (MODIS) \cite{modis} or the Visible Infrared Imaging Radiometer Suite (VIIRS) \cite{viirs}. For the satellite data, this must be processed to extract the hotspots detected by the sensors in the AOI, and converted to a format usable by WFA. The urgent decision maker reviews satellite hotspot data along with ground observations, then submit this to the VESTEC system to be fed into WFA.

A detailed local forecast of weather conditions is also required by WFA. This needs to be able to resolve the spatial and temporal evolution of the weather in the AOI. Such a detailed local forecast is required as if, for example, the AOI is a valley and so the wind direction, temperature and humidity could be greatly influenced by the topography. To obtain a local forecast, the Mesoscale Non-Hydrostatic (Meso-NH) code \cite{mesonh,mesonh_web} will be used. Meso-NH, in turn, requires one or more coarse-grained global forecasts as initial and boundary conditions, such as those produced by the European Centre for Medium-range Weather Forecasts (ECMWF) \cite{ecmwf} or the Global Forecast System (GFS) \cite{gfs}.

Finally, Wildfire Analyst requires data on the terrain in the AOI, which includes aspects such as the topography, vegetation cover, and estimated ground moisture. This data is largely static, and as such will be stored in a database which is updated infrequently. The terrain data for the AOI must be extracted from the global data and put into a format appropriate for WFA.  

At a first glance at the flowchart for the wildfire workflow (Figure \ref{wildfire_flowchart}) and the discussion above, it looks as if this workflow is easy to implement with existing WMSs, using, for example, CWL to describe the workflow. There are a number of complications with this however. The most obvious is that there are several conditional branches in the acquisition of the fire location, and with respect to the output from Wildfire Analyst. Many WMSs (and the CWL standard) do not provide this capability. This could be mitigated if we had foreknowledge of the input data that would be used for that given fire, and the type of forecast needed. Such information would allow us to produce a workflow without conditional branches descriptions for that specific case, which could be executed by the WMS. 

This approach would work initially, however more problems arise when new data arrives, for example the availability of a new global weather forecast. We would like to use this new data to obtain a more up to date local forecast and feed this into Wildfire Analyst whilst reusing its other existing inputs. This will require constructing a new workflow description for this case, such as re-calculating the weather data in combination with the existing fire position, terrain data and configuration file, then and executing it. Similarly, the urgent decision maker may decide they require a different kind of forecast. This would involve yet another workflow description with the existing inputs, but a changed configuration file to be executed. An additional complication is that we do not wish to execute a fresh instance of WFA every time new data becomes available, but instead update the existing simulation with this new data. This would require an alternative workflow step to update the running simulation rather than to execute a new one, and accordingly more workflow descriptions containing this step to be generated.

Over the lifetime of the fire there will likely be several updated forecasts produced, relying upon new data that arrives as the fire progresses. Using a traditional WMS we would need to generate several different descriptions of the workflow tailored to each scenario, and execute them separately. Whilst these workflows would be executed as separate workflows, it is important to remember that these are essentially sub-workflows of the overarching workflow for this specific fire. In order to achieve this we would need not only a WMS, but also some system that sits above the WMS to generate the specific sub-workflows and direct them to be executed as required. 

\section{Workflow Manager Design} \label{design_sec}
Based upon the discussion in Section \ref{wildfire_sec}, it is evident that existing workflow technologies do not meet our needs, and would require a software layer on top to support the execution of dynamic workflows. Rather than designing a layer which can produce CWL (or equivalent) descriptions of sub workflows for them then to be executed by an existing Workflow Management System (WMS), we decided to develop our own WMS that could execute the workflows we require natively. In this section we describe the design of our WMS, hereby the \emph{workflow manager}. 

Before we do so we must define some terminology that will be used in this section. The term \emph{workflow} is used to describe an abstract workflow of a specific type, e.g. a wildfire workflow or a space weather workflow, whilst we will use the term \emph{incident} to refer to a specific instance of a specific workflow, e.g. a wildfire workflow for a specific wildfire. Similarly, a \emph{stage} is an abstract stage of a workflow (e.g. one of the boxes in Figure \ref{wildfire_flowchart}), whilst a \emph{task} is a single execution of a stage during a workflow's execution. Note that many tasks can be associated with a single stage, as that stage may be invoked multiple times per incident, and each separate invocation is a separate task.

\subsection{Requirements} \label{requirements_sec}
Firstly it is important to define the list of requirements that the workflow system must be capable of providing, and these derive from the challenges raised in Section \ref{wildfire_sec}:
\begin{enumerate}
    \item Conditional branching\\ \label{cond_branching_req}
    The workflow manager must be able to accommodate workflows that follow different paths depending on the sources or the context of the data.
    \item Data driven\\ \label{data_driven_req}
    The workflow manager must respond to new data being pushed into it, and appropriately process this data and trigger the following workflow stages. 
    \item Persistence\\ \label{persistence_req}
    Information on the state of the workflow must be persisted so that a stage's task can take appropriate action based on previous occurrences. In the wildfire use case's \emph{Wildfire Analyst simulation} stage for example, if no WFA simulation is running, a task must execute WFA, otherwise it should update the data in a running simulation.
    \item Execution of multiple workflows simultaneously\\\label{multi_workflow_req}
    It is possible that several distinct disasters could occur simultaneously, and therefore the VESTEC system must be able to cope with running multiple incidents, possibly based upon different workflows.
    \item Parallel execution of tasks\\ \label{parallel_req}
    There can be independent workflow stages or tasks which should be executed in parallel, where possible, to speed up execution of an incident's workflow.
\end{enumerate}

\subsection{Design Considerations} 

Before establishing a design for our workflow manager, it is instructive to examine the design of other WMSs, and understand why they do not meet our requirements. The most basic components of workflows are the stages, and their relationships to each other, specified by their inputs (dependencies) and outputs. The job of a WMS is to take this information, determine the order the stages must be executed in to obtain the final goal of the workflow, create the corresponding tasks and execute them. For example, the system may construct a directional graph where the stages are the nodes of the graph and their dependencies are the edges joining the nodes. Once the graph is constructed, the manager will then traverse the graph, executing the tasks for each stage in the correct order. This method of determining the execution order of all the tasks before executing the workflow fails for our workflows, because the exact connectivity of all the stages and their corresponding tasks is not known before runtime, but is only known during execution due to Requirement~\ref{cond_branching_req}.

Our WMS must therefore not separate scheduling and running tasks into different phases of the incident's execution, but should instead have tasks themselves determine which task or tasks need to be run after they have finished. Within each task there should be logic that allows it to examine its input, from that determine what work must be completed, and then after it has finished doing its work, schedule the next task(s) to be run. Requirement \ref{cond_branching_req} is satisfied by this design, as the workflow can follow different paths depending on its input. 

\begin{figure}
    \centering
    \includegraphics{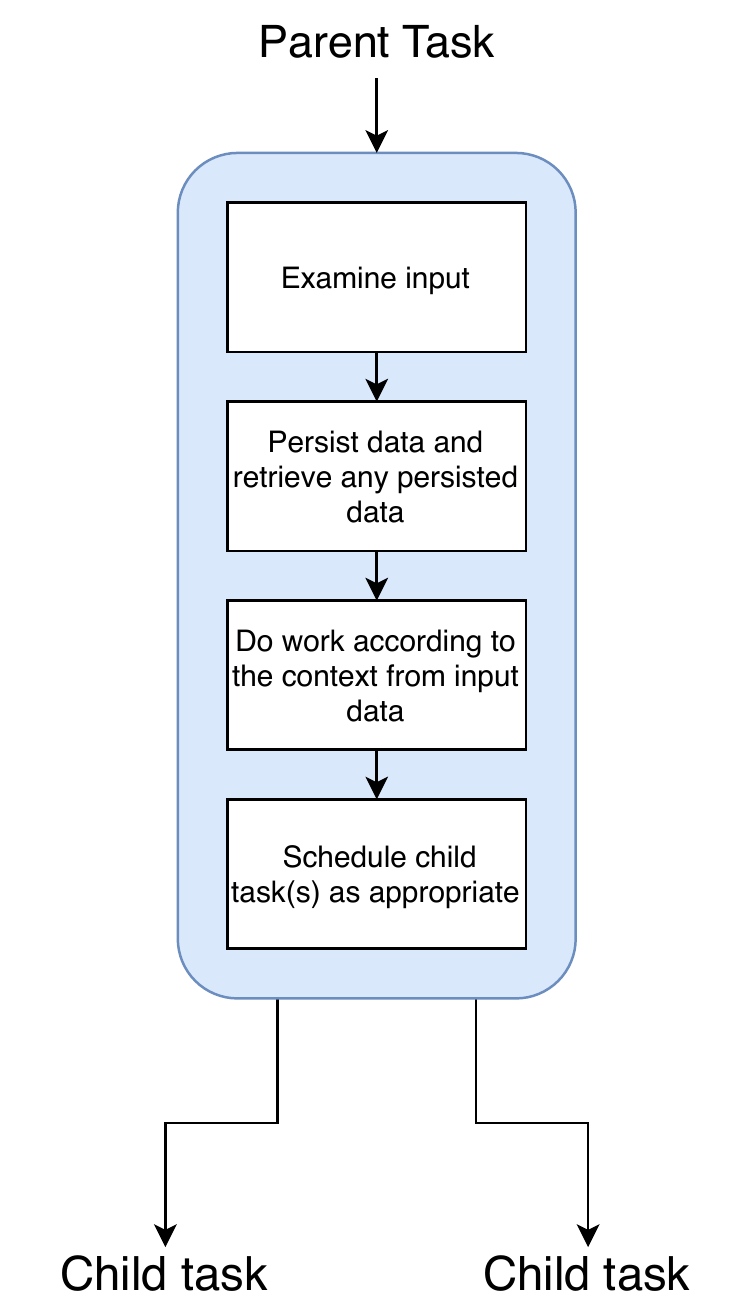}
    \caption{The anatomy of a task in our design. It is created by a parent task, examines the input it has received, optionally stores and retrieves some data for this stage, does work according to its input, then creates child tasks as required.}
    \label{task_anatomy}
\end{figure}

A difficulty with this proposed model is that each task can have only one parent task (essentially meaning that it has one dependency). This is generally not the case as in workflows, a stage can have many dependencies. Requirement \ref{persistence_req}, combined with Requirement \ref{cond_branching_req} can be used to overcome this problem. If tasks can persist some data for their stage, then they can retain a memory of what happened previously and take this into account upon execution. Consider a case of three stages: $A$, $B$ and $C$, where $C$ has $A$ and $B$ as dependencies. A task for $A$ is run, and this results in the creation of a task for $C$. $C$'s task runs, stores its input from $A$, then checks whether it has $B$'s input stored from a previous task. It does not, so it returns. Later, $B$'s task is executed, which creates a new task for $C$. This new task stores $B$'s input, then checks to see if $A$'s input has been stored previously. It has, and since it now has $A$ and $B$'s inputs, it can execute. This approach can get wasteful if a stage has many dependencies, as the stage must be called many times, once per dependency. Figure \ref{task_anatomy} displays the structure of a task in our framework.

In order to satisfy Requirement \ref{data_driven_req}, we need to have a means for the arrival of new data to trigger a new workflow task. Once this task is created, it can then trigger further tasks for the next stages as necessary to continue the workflow. This would require an interface to the workflow manager that allows external entities to request tasks be created.

Requirement \ref{multi_workflow_req} states that more than one incident should be running concurrently. To execute an additional incident we only need to create a new start task for the new incident, and providing there is an environment that supports the scheduling of multiple tasks, both incidents can execute independently from each other. If the environment contains multiple threads or processes then tasks can execute concurrently, either from different incidents, or multiple independent tasks from one incident, meeting Requirement \ref{parallel_req}. 

Parallel execution of tasks from the same incident requires that one is careful when executing tasks for stages with multiple dependencies, as the condition could arise where two tasks belonging to the same stage are running simultaneously. 
Consider the above case for the stage, $C$, with dependencies $A$ and $B$. Say that both $A$ and $B$ have completed execution, and have each scheduled $C$ to run. In principle the two instances of $C$ (one from each parent task) could be executed simultaneously on different processes. In this case, each instance would check for persisted data from previous executions, find none, because the other instance has not persisted anything yet, then persist their own input data and exit, thereby halting the workflow execution.
Problems like this can be addressed by declaring that tasks belonging to stages that take more than one input must be executed serially. 

\section{Implementation} \label{implementation_sec}
In this section we describe the implementation of our design. It is important to note that the workflow manager is just one component of many in the VESTEC system. Subsequently, it is common for tasks run by the workflow manager to invoke other components of the VESTEC system. Hence the tasks do not necessarily carry out any intensive computation, but instead coordinate the whole VESTEC system. The integration of the workflow manager with the rest of the VESTEC system will be described in Section \ref{vestec_integration}, whilst we now focus on the implementation of \emph{only} the workflow manager.

\subsection{Language}
We chose to write the workflow manager in Python version 3. Python allows us to write highly abstract code quickly, and it provides numerous modules which enable easy interfacing with other technologies. Furthermore, as mentioned above, the workflow manager's job is to coordinate the VESTEC system, and is not meant to perform any intensive work where languages such as C or C++ or Fortran may be more appropriate for performance reasons. Python also lowers the difficulty for developers writing workflow stages, as it is a well known language, and so developers are likely already familiar with it. A more in-depth discussion on the language choices for the workflow manager and VESTEC system in general can be found at \cite{vestec_system}.

\subsection{Task Scheduling}
We considered two ways of implementing task scheduling. Tasks could either directly spawn their child tasks, or they could take a message passing approach where the task sends messages to instruct a new task to be run.
The message passing architecture was chosen over a spawn style architecture. This was because a spawn-based approach imposes the risk of creating too many tasks at once, thereby adversely affecting the performance of the machine the workflow manager is running on. To implement the message passing approach, we use RabbitMQ \cite{rabbitmq}, which implements the Advanced Message Queueing Protocol (AMQP) \cite{amqp}, for handling sending and receiving messages. The Pika \cite{pika} Python module is a Python interface to RabbitMQ and used in this work. 

\begin{figure}
    \centering
    \includegraphics[width=\columnwidth]{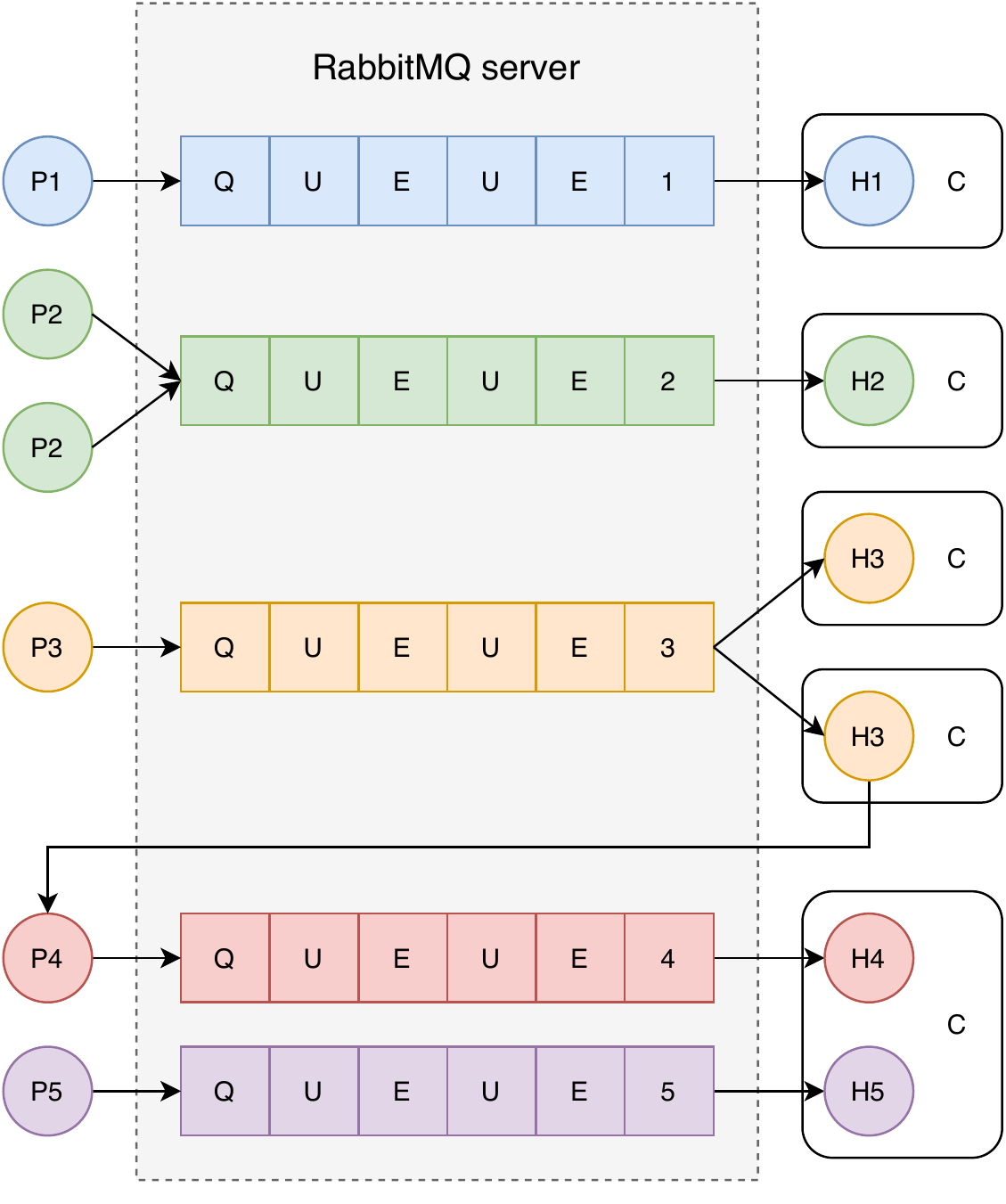}
    \caption{The various way producers (P), queues, consumer processes (C), their handlers (H) and the RabbitMQ server relate to each other. Not that a handler is able to send a message to another queue, i.e. a handler can also be a producer.}
    \label{rabbitmq_fig}
\end{figure}

RabbitMQ is a message broker and running in a separate process, acts to receive messages from a producer (the process that sends the message), and forward them to a corresponding consumer (the process that receives the message). 
Each message belongs to a named queue, and messages from a producer are stored in the appropriate queue, which are subsequently delivered to the consumer which has subscribed to that queue in a first-in first-out manner. 
Multiple producers can produce messages for a single queue, and multiple consumers can be subscribed to a single queue. In such instances the messages sent to that queue are distributed between the consumers, with each message only going to one single consumer which avoids messages being processed more than once. 
A consumer process can be subscribed to multiple queues, with the consumer providing callback functions for each it is subscribed to. These callbacks are executed upon message receipt to process the message, and it is common to refer to the queue callback function as that queue's \emph{handler}. 
It is also possible for a consumer to produce messages, effectively meaning a process can be a producer, consumer, or both. This flexibility means that a handler can generate new messages to send to other queues as required.
Figure \ref{rabbitmq_fig} displays a graphical representation of the different way producers, consumers, handlers and queues can relate to each other.

The ability for handlers to act as producers to other queues is helpful, as it permits such handlers to be chained together in sequence, much like the chaining of workflow stages together. Based upon this model, each workflow stage is represented by a handler function, with a queue for each stage. When a message from a stage's queue is delivered to the handler it is executed, which involves the handler inspecting the message, performing any necessary work, and then creating and sending messages onto other queues as required. The running handler function is the instance of that stage, e.g. a task. This execution model is illustrated in Figure \ref{task_anatomy}.
In order to reduce message size, the messages will contain the path to any data, or a handle to where it is contained rather than the data itself. Handlers are then able to access this data as required.

\subsection{Concurrency}
Requirements \ref{multi_workflow_req} and \ref{parallel_req}, as described in Section \ref{requirements_sec}, relate to the need for concurrency between different workflows or incidents, and concurrency of independent tasks within an incident. Based upon the RabbitMQ implementation, there are two different ways to achieve this. Firstly, one could provide a separate consumer process for each queue, and thus workflow stage. Secondly, a smaller number of consumers can have registered handlers for all the queues.

In the first approach, concurrency is achieved between workflow stages, as each one is handled by a separate process so each can run concurrently. This simple consumer process to workflow stage mapping meets both requirements, however will not scale to many workflows which contain many stages, as a process will be needed for each stage. Furthermore, for much of the time these processes will be idle, waiting for messages to consume. The first approach is thus a waste of resources, which is a major disadvantage. One advantage of this approach however is that because there is one consumer per stage, messages processed by each stage are processed serially, and therefore there is no risk of undefined behaviour from a stage with multiple inputs having multiple tasks execute simultaneously.

The second approach resembles a task farm, whereby a pool of worker processes (the consumers) fetch available work from the queues which they then process. Because each consumer is associated with every queue (and therefore workflow stage), provided that there are sufficient messages in the queues, then all workers can be busy consuming messages concurrently. Effectively this avoid the issue of idle processes, which was a major disadvantage of the first approach. This task farm style design also enables us to scale the number of workers according to how much work must be performed, allowing more effective management of resources. This second approach is therefore the design we have chosen, primarily due to its higher flexibility.

Care needs to be taken with this approach however, as we could enter a condition where a stage with multiple inputs is running concurrently on different processes. Therefore a mechanism that ensures only one instance of a handler is being executed at once must be implemented. One way to do this is to only register such handlers with one worker process, as then only that handler can run in serial. This workaround however can cause a performance bottleneck, because if the host process is busy processing another task, then the multi-input handler cannot be run until the first task is completed, possibly holding the workflow execution back. Another option is to flag the handlers for such stages as only being allowed to execute one at a time. If a consumer receives a message belonging to a queue from these stages, it first checks whether another consumer is executing a message from this queue. If there is another instance being executed, then the consumer re-queues this message. The implementation of such support is similar to a semaphore lock, where a consumer marks a queue as locked when it is executing the queue, and all other consumers cannot consume messages from this queue until the queue is unlocked.

The final consideration is the situation where multiple incidences of a workflow (e.g. incidents) are active at once. In this case the same set of queues, which relate to stages of the given workflow, are being used for multiple incidents simultaneously. We therefore need to be able to mark messages as belonging to specific incidents to prevent state becoming mixed up between incidents. This is fairly trivial to support by assigning each incident a unique identifier when it is created, and reference this in every message that relates to that incident.

\subsection{Persistence}
A database is used to store the state of the workflow. This both records the status of messages (e.g. sent, delivered, processing, completed, error) allowing us to track the progress of the workflow, and also allows for the persistence of workflow stage data, as per Requirement \ref{persistence_req} from Section \ref{requirements_sec}. As mentioned previously, messages contain handles to any data rather than the data itself, and therefore this database is unlikely to grow large in size due to the stored messages or persisted data. 

A timestamp is stored in the database every time a message is sent and is completed, which allows to derive statistics on the workflow's performance. These statistics can, for example, be used to identify steps in the workflow that act as a bottleneck and might need improvements. This is of importance in particular in the use case of urgent decision making, where it is desired to get results as early as possible, yielding the need of optimal performance of the workflow.

\begin{figure}
\begin{lstlisting}[language=Python]
import workflow

#defines a critical workflow stage
@workflow.critical
@workflow.handler
def A_handler(message):
    #get the incident this message belongs to
    IncidentID = message["IncidentID"]
    
    #persist some data
    workflow.Persist.Put(Incident = incidentID, data = {"key": "value"})
    
    #get persisted data as a list
    data = workflow.Persist.Get(incident = IncidentID)
    
    #send a message to "B_queue"
    workflow.Send(queue = "B_queue", message = message)
    
@workflow.handler
def B_handler(message):
    IncidentID = message["IncidentID"]
    
    workflow.Complete(incident = IncidentID)
    
#Open connection to RabbitMQ
workflow.OpenConnection()

#associate the handlers with their queues
workflow.RegisterHandler(handler = A_handler, queue = "A_queue")
workflow.RegisterHandler(handler = B_handler, queue = "B_queue")

#create an incident
IncidentID = workflow.CreateIncident(name="some_name",kind="some_description")
    
#create a dictionary for the message and include the IncidentID
message={ "IncidentID" : IncidentID}

#enqueue message to A to start workflow
workflow.send(queue = "A_queue",message = message)

#send the enqueued message
workflow.FlushMessages()

#execute the workflow (this starts up the consumer in an infinite loop waiting for new messages)
workflow.execute()
\end{lstlisting}
\caption{Example code to demonstrate the API for implementing workflows with our workflow manager. The code defines some handlers, registers them to queues, creates a new incident, sends the initial message to start the workflow, then executes the workflow.}
\label{code_fig}
\end{figure}

\subsection{API} \label{api_sec}
When designing the interface to the workflow manager, we aimed for a design that was simple for workflows to be integrated. We therefore wanted to abstract away as much of the inner workings of the workflow manager as possible, so as to allow the programmer to focus on implementation of their workflow. To this end, the workflow manager presents a number of simple functions and decorators for the implementer to use.

Firstly, handler functions are identified as handlers through a decorator, and these functions accept only a single argument, the message, which is a Python dictionary. The decorator itself logs receipt of the message, unpacks the message into a python dictionary, checks that the message belongs to a valid incident, executes the handler, logs successful completion of the handler, and lastly sends any messages generated by the handler onto their queues. It also catches any exceptions thrown by the handler and logs this. At present, if the handler throws an exception, this incident is marked as being in "error", and no new messages from this incident are allowed to run. This behaviour was chosen as the workflow has no way of dealing with the outcomes of a failed stage. We could in principle (if the exception is not fatal), re-queue this message to be handled at a later time, however this is not implemented as of the time of writing.

The handler can also be decorated with a \emph{critical} decorator, which denotes it as being a handler which can only be run once at a time, such as a handler for a multi-input stage. This decorator checks if another consumer is already executing a message from this queue. If it isn't, it sets a flag in the database to say that it is processing a message from the queue, processes the message, then unsets the flag. If another consumer has set the flag, the consumer re-queues the message as it cannot be processed at that moment.

Functions for enqueueing messages to be sent, and for storing and retrieving persisted data are available to be called within a handler. The send function accepts the name of the queue which the message is to be sent to, and the contents of the message (as a Python dictionary). However, this function only enqueues messages and does not physically send them. Instead, messages are sent within the handler's decorator function only after the handler has exited cleanly. This prevents messages being sent from a handler which subsequently fails, possibly leading to undefined behaviour further down the workflow. 

Persistence of workflow stage data is implemented by allowing each stage to write an entry containing the incident ID, queue name, and a reference to necessary data stored as a JSON to a database table. Each stage can also look up all the persisted data for that stage and Incident ID, returning a list (ordered by time) of the persisted data from each subsequent execution of the handler.

A number of miscellaneous functions are also provided, for instance permitting the creation of an incident, which registers a new incident with the database and returns the ID of this incident for use in a message. Functions to mark an incident as completed, as well as to cancel a running workflow are also provided. These finalisation functions modify the state of that incident to completed or cancelled, then enqueue a special cleanup message which only runs after all other messages belonging to that incident have been processed. This cleanup function deletes any persisted information for workflow stages for that incident, and deletes any locks associated with the incident. 

Figure \ref{code_fig} demonstrates a basic example of the API for writing a workflow with the workflow manager. This simple workflow consists of two steps, $A$ and $B$, where $A \to B$. The handlers demonstrate the minimum code required to make a critical handler, persist and retrieve persisted data, and send messages. This example also demonstrates creating an incident, sending the initial message to the workflow, and executing the consumer process. In reality the consumer, definition of handlers and workflow creation would be in separate Python Files.

\section{Integration with the VESTEC system} \label{vestec_integration}
As mentioned in Section \ref{implementation_sec}, the workflow manager is primarily used to coordinate the VESTEC system to carry out the appropriate tasks to aid in urgent decision making. Its role is to invoke the other components of the VESTEC system to complete the necessary work. In this section we describe how the workflow manager integrates with the other components of the system. The design of the system is presented in detail in \cite{vestec_system}.

\subsection{Data Manager}
As the VESTEC system uses HPC machines to run the large simulations, it may have data distributed over many machines. We therefore have a component called the Data Manager (DM) that tracks the location of the data, and is responsible for moving this from machine to machine as necessary, or for uploading/downloading from the web interface or from external data sources. In a workflow stage where data needs to be downloaded from the internet, or moved to/from a HPC machine, the workflow stage's handler instructs the DM to action this. Each data item has a unique identifier associated with it, and it is this identifier that is contained within messages between workflow stages, rather than the data itself.

\subsection{HPC Interface}
\begin{figure}
    \centering
    \includegraphics[width=\columnwidth]{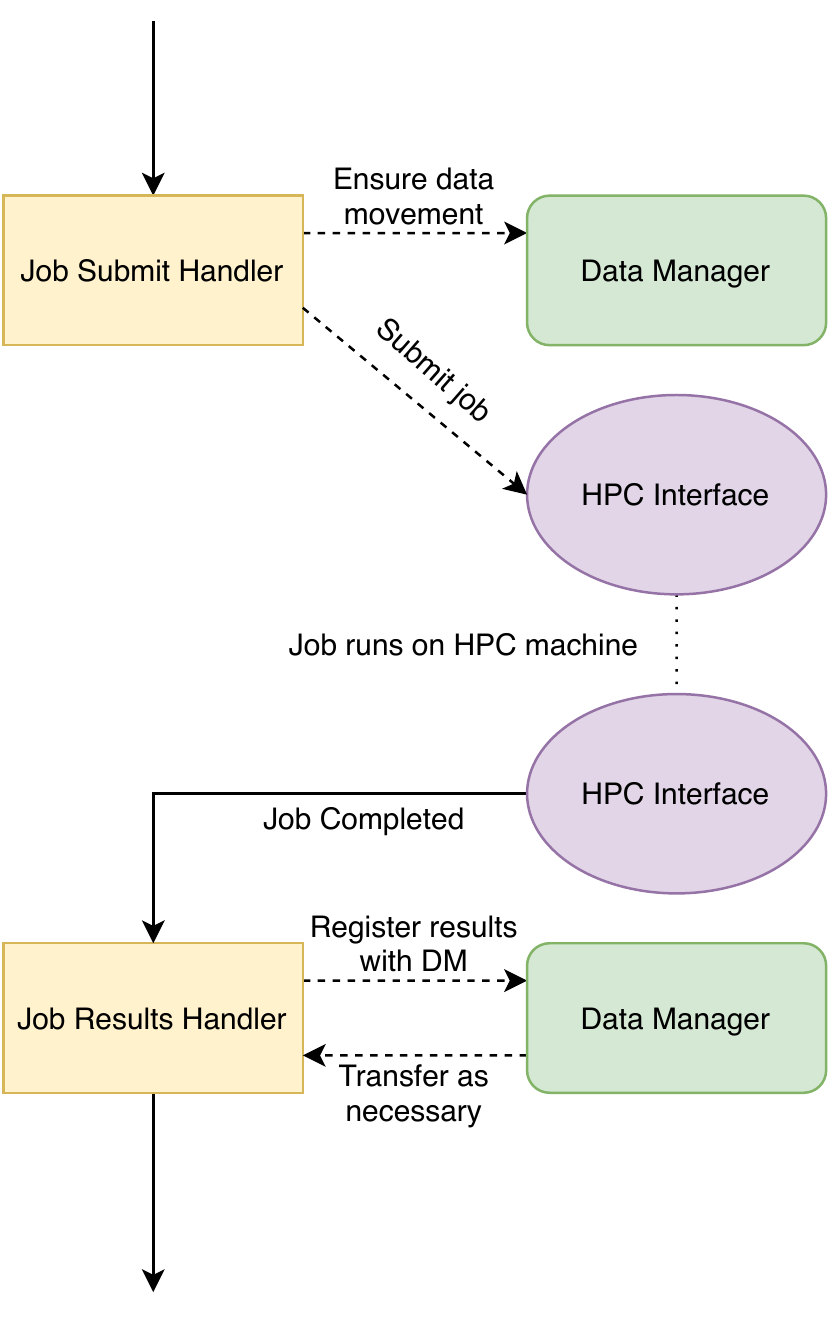}
    \caption{How the workflow submits and retrieves the results of a job run on a HPC machine with use of the workflow handlers (orange), Data Manager (green) and the HPC interface (mauve).}
    \label{hpc_interface_fig}
\end{figure}

The VESTEC system must be able to submit jobs onto HPC machines, as well as check on their status, and create the files necessary to run a simulation on the machine. The HPC interface is responsible for this. A handler for submitting a job to a HPC machine would therefore instruct the DM to ensure all the files needed for the job are placed into the correct location on the machine, then instruct the HPC Interface to launch this job, instructing it which workflow queue to notify upon completion of the job. Once this has occurred, the handler exits, allowing other queued handlers to be executed, preventing that consumer from blocking until completion of the simulation. Upon job completion, the HPC Interface sends a message to the appropriate queue to notify it of job completion. The job completion handler then registers the output file with the DM, and possibly retrieves the results if appropriate, using the DM. An example of this is shown in Figure \ref{hpc_interface_fig}. A similar procedure works for simulations such as Wildfire Analyst which remain executing in the background waiting for new data. In this case, the HPC interface sends a message to the appropriate queue when a new forecast file is produced, rather than upon simulation completion.

\subsection{External Data Interface}
Another key requirement of the VESTEC system is that it must be able to obtain new data as the incident is ongoing, and then feed this data into the appropriate workflow stage to update the predictions. The External Data Interface (EDI) is the component responsible for this. It allows for data to be pushed to a pre-defined endpoint on the VESTEC system from outside, e.g. from the urgent decision maker via the Web Interface, or by a sensor which has knowledge of the specific URI associated with this endpoint. This action triggers a message to be sent to a workflow stage to handle this data. The EDI can also be used to pull data, whereby it periodically polls a web endpoint (e.g. weather forecasts or satellite data websites) and if new data is available, it sends a message containing the web endpoint's header to the appropriate workflow stage informing it of the new data so it can handle this. That workflow stage then can instruct the DM to download the data if it deems this necessary. 

\subsection{User Interfaces}
The urgent decision maker interacts with the VESTEC system via a use-case specific user interface. The wildfire use-case described in this paper involves a bespoke GUI that is already part of Tecnosylva's suite of tools. Furthermore, the VESTEC system also provides a web interface which is designed to enable the management, auditing and debugging of incidents, and amongst other functionality displays a graph of the actual tasks that have been executed during an active incident and, therefore, visualizes the actual conditional branches that have already been executed. There are also performance metrics reported, for insance this workflow graph also includes the runtime of each tasks, displaying the the performance data that are continuously collected during the execution.

There were two strategies adopted to support the integration of numerous third-party GUIs and also the VESTEC web-interface. Firstly, a general purpose RESTful API was developed which is designed to be applicable to any use-case and primarily exposes management functionality. Furthermore, use-case specific interaction endpoints can be exposed and this is provided by the interaction between the EDI and workflow system. When workflows initialise for an incident, they can create any number of EDI endpoints, and hence the GUIs associated with a specific use-case can push appropriate information into the VESTEC system

\begin{figure*}
    \centering
    \includegraphics[width=\textwidth]{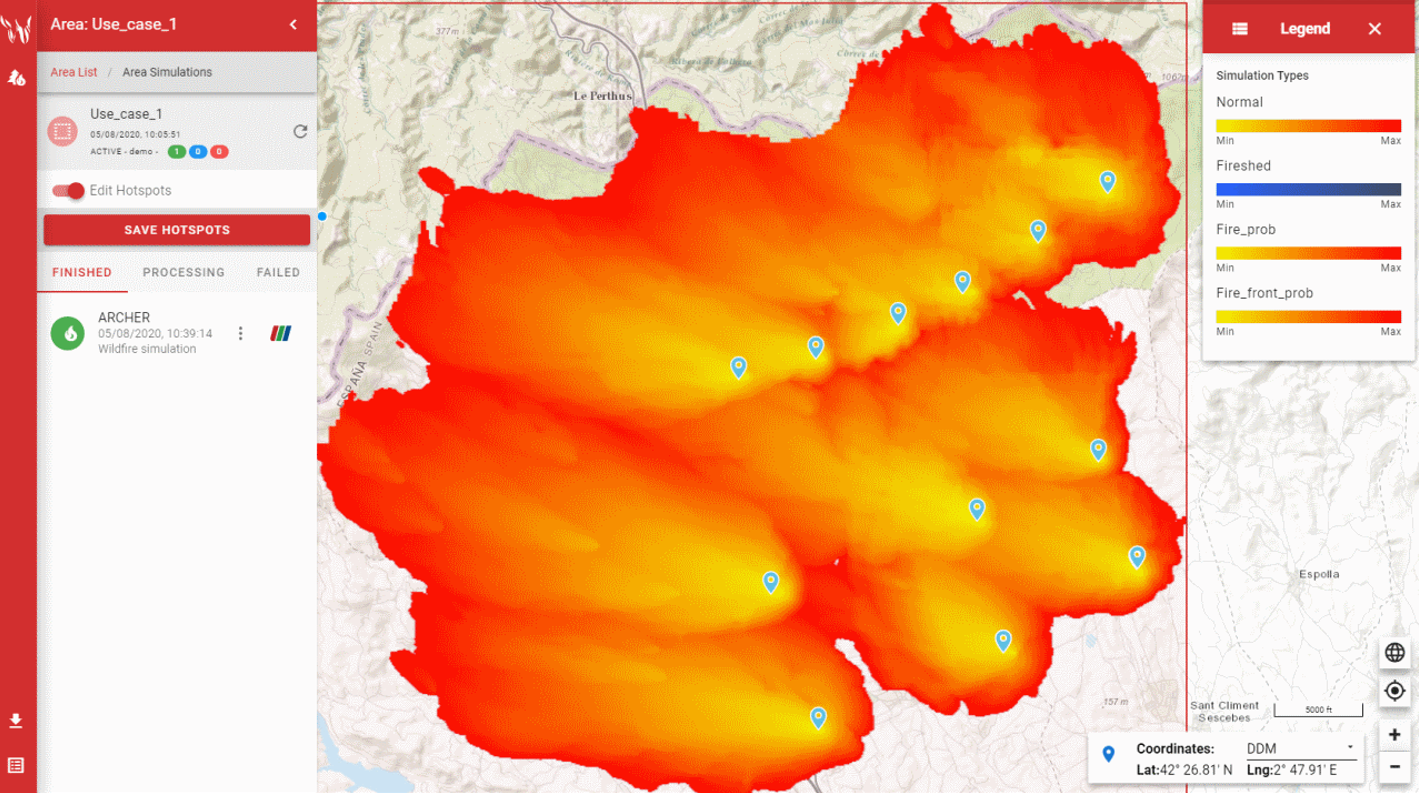}
    \caption{WildFire Analyst GUI illustrating the progression of the La Jonquera fire from 2012. The screenshot shows the initial hotspots (blue markers) and the predicted fire progression from a simulation on the ARCHER supercomputer.}
    \label{wfa_fig}
\end{figure*}

Figure \ref{wfa_fig} illustrates the GUI of WFA and, as a third party GUI, illustrates an examples of an external system interacting with the VESTEC system. This enables users to automatically consume hotspot data which originates from satellites, tweak the location and properties of resultant hotspots in the landscape, drive simulations of fire progression, and display results. These all rely on a combination of the system's general purpose RESTful API, and use-case specific services exposed via the EDI, and it demonstrates that a rich set of functionality can be supported based upon the design decisions adopted.

\section{Conclusion} \label{conclusion_sec}

In this paper we have presented a Workflow Management System (WMS), the \emph{workflow manager}, that we created to execute the Urgent Computing workflows required by the VESTEC project. We initially described a workflow required for fighting wildfires, and from this demonstrated why existing workflow technologies are not suitable for executing these sorts of workflows. We then generated a list of requirements that our WMS needed to meet and developed the most suitable general design to address the requirements. We then outlined the implementation of the workflow manager, and finally described how it integrated with the rest of the VESTEC system in order to execute an Urgent Computing workflow.  

One of the most notable feature of our workflow manager is that it enables data-driven workflows, whereby new data can arrive and alter the flow of the workflow during execution. This is crucial for the needs of Urgent Computing, where the workflow must adapt to the changing situation of a disaster as it is progressing. Our design is in contrast to most existing workflow technologies, which require a complete description of the steps and dependencies before execution of the workflow can proceed, thereby precluding the ability to alter the workflow mid-execution.
This functionality is achieved through our approach of defining each workflow stage as an entity that is triggered by an incoming message. These entities processes the message, and then can send successive messages onto other stages based upon the context as required, which then triggers these subsequent stages. New data can be sent into the workflow by sending a message to the appropriate stage to handle this, and each stage itself determines who to send the next messages to, so there is no need for constructing the full workflow execution graph before execution as the workflow does this at runtime.

One limitation with our design is that each running instantiation of a workflow stage can only accept a single input, however many workflow stages in real life can require multiple inputs. Our solution to this problem is for multi-input stages to take their input, and then store it. These stages then check against previously stored inputs to track whether all the prerequisite data is present, if not then it merely exits, whereas if so then the stage executes. This approach can get wasteful if a stage has many dependencies, as the stage must be called many times, once per dependency, however for the workflows we are dealing with there are no more than a handful of inputs to each stage. This does however make the workflow manager less suited for workflows containing stages with many (e.g. $>100$) dependencies.

The workflow manager was implemented in Python, using RabbitMQ to manage the messaging. There are a group of consumer processes that consume messages from the various workflow stages' queues to execute the workflow. There is also a database to store information required for the workflow stages, and to record the workflow's progress. The way in which we have constructed the workflow manager, means that we are able to scale up or down the number of consumers (workers) according to the computational load. This is beneficial if there are many incidents occurring simultaneously, and/or more concurrency in workflow stage executions is required. Additionally, a feature of RabbitMQ is that messages are persisted in the queues, so even if the VESTEC system fails, the workflow's progress is not lost, and it can be restarted upon the system being brought back up, although workflow that were executing stages will have to start from the beginning. For these sorts of disaster response use-cases, this resiliency is highly desirable because it means that a failure of the VESTEC system does not lead to progress being lost with the workflow.

The API for the workflow manager makes it relatively easy for third parties to implement a workflow themselves. This means that workflows can be quickly and clearly written, allowing the VESTEC system to be extended to many more use cases in the future, which is a key aim of the project. The modular nature of the VESTEC system also allows each component to have a well defined functionality, and allows the workflow programmer to use them as necessary to conduct their urgent computing workflow.

At present we have fully implemented the Wildfire workflow, such that it can pull new data from satellites and weather forecasts, run simulations and send results back to the urgent decision maker. Tecnosylva have integrated the VESTEC system's Web interface APIs into their Wildfire Analyst client GUI, so that the VESTEC system can be accessed directly by urgent decision makers using their software, as was shown in Figure \ref{wfa_fig}. In terms of further work, two other use cases, which tackle space weather and mosquito borne diseases, are currently being integrated into the VESTEC system. Whilst these workflows are drastically different in terms of their application, the structure of their workflows are similar to the wildfire case covered in this paper. 
In the further development of the VESTEC system itself, which was described in Section~\ref{vestec_integration}, it is our aim to exploit the performance statistics that are collected by the workflow manager to, for example, optimize the choice of where the HPC interface runs a simulation in a specific workflow step.

\section*{Acknowledgments}

The research leading to these results has received funding from the Horizon 2020 Programme under grant agreement No.\ 800904. This work used the ARCHER UK National Supercomputing Service (http://www.archer.ac.uk).


\bibliographystyle{IEEEtran}
\bibliography{IEEEabrv,workflows.bib}

\end{document}